\documentclass[fleqn,usenatbib]{mnras}

\usepackage{newtxtext,newtxmath}
 \usepackage{booktabs}
 \usepackage{graphicx}

\usepackage[T1]{fontenc}

\DeclareRobustCommand{\VAN}[3]{#2}
\let\VANthebibliography\thebibliography
\def\thebibliography{\DeclareRobustCommand{\VAN}[3]{##3}\VANthebibliography}

\usepackage{graphicx}	
\usepackage{amsmath}	
\usepackage{hyperref}
\usepackage{subcaption}
\hypersetup{unicode=true, colorlinks=true, linkcolor=[rgb]{0.53, 0.15, 0.34}, citecolor=[rgb]{0.0, 0.27, 0.42}, filecolor=[rgb]{1.0, 0.13, 0.32}, urlcolor=[rgb]{0.53, 0.15, 0.34}}
\newcommand{\nustar}{\textit{NuSTAR}~}
\newcommand{\swift}{\textit{Swift}/BAT~}
\newcommand{\xmm}{XMM--\textit{Newton}~}

\usepackage{color}
\usepackage{array}
\usepackage[dvipsnames]{xcolor}

\title[Achromatic rapid X-ray flares in LS~I~+61~303]{Achromatic rapid flares in hard X-rays in the $\gamma$-ray binary LS~I~+61~303}

\author[Saavedra et al.]{
Enzo A. Saavedra$^{1}$\thanks{e-mail: enzosaave@fcaglp.unlp.edu.ar},
Gustavo E. Romero$^{1,2}$, 
Valenti Bosch-Ramon$^{3}$,
Elina Kefala$^{3}$.
\\
$^{1}$Facultad de Ciencias Astron\'omicas y Geof\'{\i}sicas, Universidad Nacional de La Plata, Paseo del Bosque, B1900FWA La Plata, Argentina \\
$^{2}$Instituto Argentino de Radioastronom\'ia (CCT La Plata, CONICET; CICPBA; UNLP), C.C.5, (1894) Villa Elisa, Buenos Aires, Argentina \\
$^{3}$Departament de F\'{i}sica Qu\`antica i Astrof\'{i}sica, Institut de Ci\`encies del Cosmos (ICC), Universitat de Barcelona (IEEC-UB), \\ Mart\'{i} i Franqu\`es 1, E08028 Barcelona, Spain.
}

\date{Accepted XXX. Received YYY; in original form ZZZ}

\pubyear{2023}

\begin{document}
\label{firstpage}
\pagerange{\pageref{firstpage}--\pageref{lastpage}}
\maketitle

\begin{abstract}
We report on the presence of very rapid hard X-ray variability in the $\gamma$-ray binary LS~I~+61~303. The results were obtained by analysing \nustar data, which show two achromatic strong flares on ks time-scales before apastron. The \swift orbital X-ray light curve is also presented, and the \nustar data are put in the context of the system orbit. The spectrum and estimated physical conditions of the emitting region indicate that the radiation is synchrotron emission from relativistic electrons, likely produced in a shocked pulsar wind. The achromaticity suggests that losses are dominated by escape or adiabatic cooling in a relativistic flow, and the overall behaviour in hard X-rays can be explained by abrupt changes in the size of the emitting region and/or its motion relative to the line of sight, with Doppler boosting potentially being a prominent effect. The rapid changes of the emitter could be the result of different situations such as quick changes in the intra-binary shock, variations in the re-accelerated shocked pulsar wind outside the binary, or strong fluctuations in the location and size of the Coriolis shock region. Although future multi-wavelength observations are needed to further constrain the physical properties of the high-energy emitter, this work already provides important insight into the complex dynamics and radiation processes in LS~I~+61~303. 
\end{abstract}



\section{Introduction}\label{S_intro}

Only a relative small number of binary systems have been observed to emit high-energy (HE; 0.1-100 GeV) and very-high-energy (VHE; $>$ 0.1 TeV) gamma rays. Despite their rarity, these systems are valuable for investigating astrophysical mechanisms of particle acceleration because of their proximity, which allows to obtain some knowledge of the varying ambient conditions and their capability to produce relativistic populations of particles evolving on short time-scales, as evidenced by their fast variability across the electromagnetic spectrum.

Despite their rarity, various different types of binaries have been identified as emitting HE radiation. These systems include at least three high-mass X-ray binaries (Cyg X-1, Cyg X-3, and SS 433), several low-mass-star/neutron-star binaries, two colliding wind binaries ($\eta$-Car and WR11), and several so-called gamma-ray binaries: systems likely formed by a pulsar and a massive star, usually of Be type. Among all these objects, with the exception of $\eta$-Car and SS 433, only gamma-ray binaries exhibit notable VHE emission \citep{Dubus2013A&ARv..21...64D, Dubus2015CRPhy..16..661D, pb19,Chernyakova2020}.

LS~I~+61~303 is an outstanding example of the group of gamma-ray binaries \citep{Albert2006Sci...312.1771A}. It is a high-mass X-ray binary system that consists of a rapidly rotating B0 Ve star of 10--15 M$_{\odot}$, according to \citet{Casares2005MNRAS.360.1105C}, and a compact object whose identity remained unknown for many years \citep[e.g.,][]{Romero2007A&A...474...15R, Romero2008IJMPD..17.1875R, BoschRamon2006A&A...459L..25B,massi09}. 
The recent detection of radio pulses, however, clearly indicates that the compact object is a rotating neutron star (NS) rather than a black hole \citep{Weng2022NatAs...6..698W}. This finding provides new insights into the system and the mechanisms operating in this and other gamma-ray binaries. 
The orbital period of LS~I~+61~303 is $P_{\rm orb} =26.496\pm 0.003$ d, the eccentricity of the orbit is $e=0.72\pm0.15$, and the periastron passage occurs at the orbital phase $\phi_{\rm orb}=0.23\pm 0.03$ \citep{Casares2005MNRAS.360.1105C, Aragona2009}. 
HI measurements indicate that the system is situated at a distance of 2.0$\pm$0.2 kpc \citep{Frail1991AJ....101.2126F}. 

Despite the past uncertainty regarding the nature of the compact object, the detection of a couple of soft gamma-ray flares in the direction of LS~I~+61~303 \citep{Dubus2008, Burrows2012GCN.12914....1B} led to speculation that the primary object in the system might not be just a pulsar but a magnetar \citep[see also][]{Torres2012ApJ...744..106T, Papitto2012ApJ...756..188P}. This hypothesis is further supported by the measurements of the period obtained with the FAST radio telescope \citep{Weng2022NatAs...6..698W}. The predictions of the magnetic-dipole braking theory suggest a polar magnetic field strength of $B_p \approx 6.4 \times 10^{19} P\dot{P}^{1/2} G \sim 7 \times 10^{14} G$, a value that is among the highest in the Galactic magnetar population \citep{Surovov2022ApJ...940..128S}.

In 1978, Gregory and Taylor first reported that LS~I~+61~303 exhibited high radio variability during a survey for highly variable radio sources in the Galactic plane. Shortly after the discovery, \citet{Taylor1982ApJ...255..210T} and \citet{Gregory_2002} found a periodic modulation of its radio emission with the orbital period of the system. The zero-phase epoch used today dates back to the first radio detection made by \citet{Gregory1978Natur.272..704G}. Additionally, \citet{Paredes1987PhDT.......113P} observed super-orbital radio variability with a modulation of the radio outburst peak occurring approximately every 4 yr. \citet{Peracaula1997PhDT.......265P} detected small-amplitude radio variability on time-scales as short as 1.4 h during the flux decay immediately following the expected periodic radio outburst.
The soft X-ray emission of LS~I~+61~303 is also modulated by the orbit and exhibits outbursts between the orbital phases 0.4 and 0.8 \citep{Li2012ApJ...744L..13L, Abdo2009ApJ...701L.123A, Zamanov2014A&A...561L...2Z}. The MAGIC collaboration has reported correlated X-ray and VHE gamma-ray emission from the source during $\sim$ 60 per cent of one orbit, suggesting a single particle population as the origin of the emission \citep{mag2009,Zabalza2011A&A...527A...9Z}.

The orbital modulation of LS~I~+61~303 is also observed at other wavelengths. \citet{Paredes1986A&A...154L..30P} initially reported optical variability, which was later found by \citet{Mendelson1989MNRAS.239..733M} to follow the orbital period. \citet{Paredes1997A&A...320L..25P} observed a similar pattern in X-rays. At X-rays, the source was first identified using the Einstein satellite \citep{Bignami1981ApJ...247L..85B} and has since been monitored at various energies with different instruments, including {\it ROSAT}, {\it ASCA}, {\it RXTE}, {\it XMM-Newton}, {\it INTEGRAL}, {\it Swift/XRT}, and {\it Chandra} \citep{Sidoli2006A&A...459..901S, Paredes2007ApJ...664L..39Pchandra, Torres2012ApJ...744..106T, Chernyakova2012ApJ...747L..29C}. The X-ray flux was found to fluctuate over time-scales of days \citep{Goldoni1995A&A...299..751G} or even shorter \citep{Li2012ApJ...744L..13L}. {\it Fermi-LAT} detected the emission of the source at HE gamma rays. This emission consists of periodic outbursts occurring shortly after periastron passage ($\phi_{\rm orb} \sim$ 0.3--0.45; \citealt{Hadasch2012ApJ...749...54H}), so it appears before the VHE radiation detected by Cherenkov telescopes.

\citet{Mestre2022A&A...662A..27M} found that LS~I~+61~303 exhibits optical micro-flares with time-scales of one day and that these micro-flares are correlated with gamma-ray emission. Rapid fluctuations in the emission have been frequently associated with the possible presence of a magnetar \citep{Torres2012ApJ...744..106T, Zamanov2014A&A...561L...2Z, Suvorov2022ApJ...940..128S}.

Regarding the X-ray energy range, the source was observed with {\it Swift} \citep{Esposito2007A&A, Acciari2009ApJ, Dai2016MNRAS.456.1955D, Chernyakova2017MNRAS.470.1718C} and {\it INTEGRAL} \citep{Chernyakova2006MNRAS,Zhang2010MNRAS,Li2014ApJ} satellites. {\it INTEGRAL} data in the range 20--60 keV show a hard spectrum with index $\Gamma\sim 1.5$ when the compact object is around apastron \citep{Chernyakova2006MNRAS}, with fluxes in the range $\sim 2-4 \times 10^{-11}$ erg s$^{-1}$ cm$^{-2}$. Around phases 0.4--0.6, the spectrum is somewhat softer with values of $1.7 \pm 0.4$ \citep{Chernyakova2006MNRAS} or $1.9\pm 0.2$ \citep{Zhang2010MNRAS}, with similar fluxes. For the phase range 0.0--0.4, which includes the periastron passage, \cite{Li2014ApJ} found an X-ray photon index in the 18--60 keV band of $1.9^{+0.47}_{-0.40}$, with a flux of $1.6\pm 0.3\times 10^{-11}$ erg s$^{-1}$ cm$^{-2}$. In the case of {\it Swift}/XRT, a cumulative spectrum through five orbits was obtained by \cite{Esposito2007A&A}. It is well described by an absorbed power-law model with an average index $\Gamma=1.78\pm 0.05$. \cite{Acciari2009ApJ} reported observations with both {\it Swift} and {\it RXTE} with a variable X-ray flux in the 2006/2007 season, in the range $\sim 0.5-3.0 \times 10^{-11}$ erg cm$^{-2}$ s$^{-1}$ over the orbit, with a variable spectrum. These data are restricted to the energy band of 2--10 keV.

 In addition to the above mentioned studies in the hard X-ray band, \citet{Chernyakova2006MNRAS} conducted an extensive investigation at lower energies using multiple \xmm observations. Throughout these observations, they consistently found an index $\Gamma~\sim~1.5$ in the phase range 0.2--1, regardless of whether the flux was high or low.

 \citet{Sharma2021MNRAS.500.4166S} analysed the variability of radio and X-ray emission through simultaneous observations. In radio, they used the {\it AMI Large Array Telescope} in the frequency ranges 13-15.5 GHz and 15.5-18 GHz. The X-ray emission was observed by \xmm in the energy range of 0.3-10 keV. The observations focused on the orbital phase range 0.696--0.711. 
They found that the radio and X-ray emission are correlated up to 40 ${{\ \rm per\ cent}}$ once the long term trends are removed.

\citet{Rea2010MNRAS} performed an observation using the \textit{Chandra} telescope in the energy band of 0.3--8 keV. The observation was centered on phases 0.94--0.98 and unveiled a moderate spectral variation ($\Gamma \sim 1.70-1.83$). Two minor flares (with $\Gamma \sim 1.70\pm0.02$) were also detected. The objective of these observations was to detect X-ray pulsations, but no periodic signals were found in the observed frequency range. The low X-ray pulsed fraction ($\lesssim 10 \%$) suggests that the X-ray emission likely originates from the inter-wind shock or inner-pulsar wind zone.

\citet{Lopez2023} recently analysed the {\it RXTE} historical archive to determine the orbital period, which they found to be 26.6 $\pm$ 0.3 days, consistent with what was found previously. They also conducted a timing study but found no evidence of a possible periodic pulse at X-rays.

Very rapid variability was reported by \cite{Smith2009ApJ} using {\it RXTE} in the energy band of 3--10 keV. They detected an exceptionally large X-ray flare, with a peak flux of $7.2 \times 10^{-11}$ erg cm$^{-2}$ s$^{-1}$ and a duration of $\sim 100$ s. During this time, the flux increased by a factor of 6, which was (aside from the more controversial magnetar-like activity mentioned above) the strongest X-ray flare observed so far in this source. The event occurred toward the apastron, with an integrated luminosity, assuming a distance of 2 kpc, of $\sim 3.4 \times 10^{34}$ erg s$^{-1}$. The spectral index during this flare was hard: $\Gamma=1.4\pm 0.1$. Other two smaller flares were also observed soon afterwards.

In this paper, we investigate the hard X-ray behaviour of the source using recent \swift and {\it NuSTAR} data. We found the usual orbital variability in the \swift data and discovered two fast flares before apastron passage with {\it NuSTAR}, each presenting internal structure and typical variability time-scales of $\sim 1$~ks with a total duration of $\sim 10$~ks. These flares are the most prominent reported at hard X-rays (3--79 keV). They occurred in similar phases as the flares found by \cite{Smith2009ApJ} with {\it RXTE} at lower energies but with a softer spectrum. The spectrum found here is, however, compatible with that derived from {\it INTEGRAL} data, although with somewhat lower fluxes.
 
The article is structured as follows. \hyperref[sec:data]{Section~\ref{sec:data}} describes the data set and the corresponding analysis. The results are presented in \hyperref[sec:results]{Section~\ref{sec:results}}. We then provide a discussion on the interpretation of these rapid flares and the possible underlying physical mechanisms in \hyperref[sec:Discussion]{Section~\ref{sec:Discussion}}. In the final section (\hyperref[sec:Conclusiones]{Section~\ref{sec:Conclusiones}}), we offer our conclusions.

\section{Observations and Data Analysis} \label{sec:data}

\subsection{{\textit {\textbf Swift}}/BAT data}

NASA launched the {\it Swift} satellite in 2004, which includes the {\sl Swift Burst Alert Telescope} (BAT) as part of its multi-wavelength observatory \citep{Gehrels2004ApJ...611.1005G}. BAT uses a coded aperture mask to detect and locate gamma-ray bursts (GRBs) in the 15--150 keV energy range. Its wide field of view and high sensitivity allow it to provide initial positions for follow-up observations by other instruments on the {\it Swift} satellite and ground-based telescopes in less than 20 s.

BAT has several technical capabilities that make it useful for a range of astronomical studies. Its large field of view allows it to survey a significant fraction of the sky each day, which is useful for studies of the cosmic X-ray background and the large-scale structure of the universe. Additionally, its high angular resolution enables it to accurately locate and study a variety of astronomical objects, such as active galactic nuclei, X-ray binaries, and supernova remnants \citep{Barthelmy2005SSRv..120..143B}.
\begin{figure}
    \centering
    \includegraphics[width=0.45\textwidth]{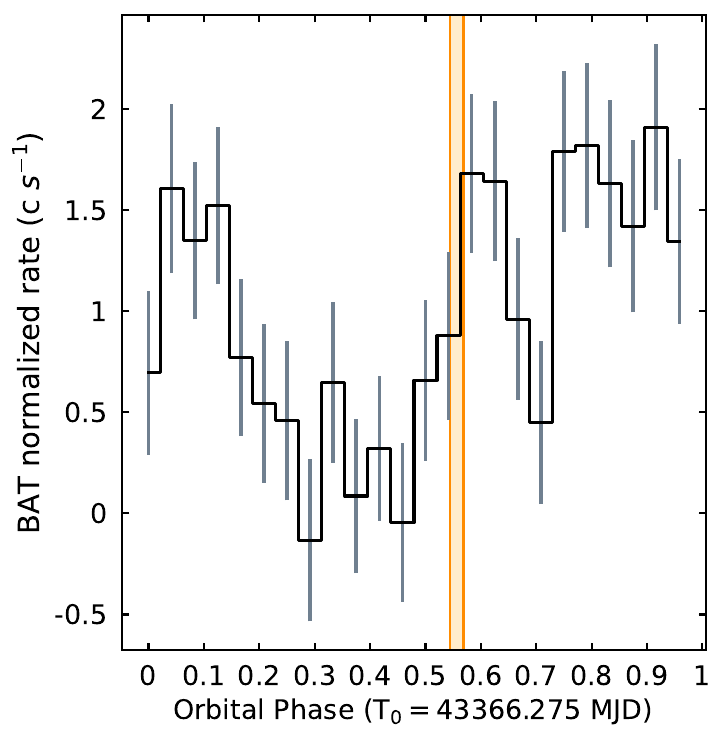}
    \caption{{\it Swift}/BAT folded light curve using 24 bins, an orbital period of 26.496 days, and the reference epoch 43366.275 MJD \citep{Gregory_2002}. The {\it NuSTAR} observation of LS~I~+61~303 analysed in this work occurred at the orbital phase indicated by the orange stripe, spanning $\sim$2.4 per cent of the orbital period, ranging from 0.544 to 0.568 in orbital phase. Note that the apastron passage takes place at phase 
    $\sim 0.73$ \citep{Casares2005MNRAS.360.1105C} (or at phase $\sim 0.775$ when adopting the results from \citealt{Aragona2009}).}
    \label{fig:bat}
\end{figure}

We investigated the full orbital light curve of LS~I~+61~303 available on \swift service up to 2022 November 16. This public website\footnote{\href{https://swift.gsfc.nasa.gov/results/transients/weak/LSIp61303/}{ LS~I~+61~303}} offers over 1000 light curves of hard X-ray sources, spanning over 9 yr. In \hyperref[fig:bat]{Fig.~\ref{fig:bat}}, we present a light curve that was constructed using 24 bins for the whole orbit of 26.496 d, at the reference epoch 43366.275 MJD. In the present work, we mostly focus on the period of the {\it NuSTAR} observation of LS~I~+61~303, which was conducted before apastron passage and covered approximately 2.4 per cent of the orbit. This observation spanned from 0.54--0.57 in orbital phase, as indicated by the orange stripe in the figure. It should be noted that the apastron passage of LS~I~+61~303 should occur within the phase range 0.73--0.775 \citep{Casares2005MNRAS.360.1105C,Aragona2009}. The \swift light curve shows a significant decrease in flux during the periastron passage, while in the region near the apastron, the source presents a flux several times higher. This pattern agrees with the findings of \citet{Esposito2007A&A}, who also analysed {\it Swift}/XRT data.

\subsection{{\textit {\textbf NuSTAR}} data}

The {{\sl Nuclear Spectroscopic Telescope Array}} (\textit{NuSTAR}; \citealt{Harrison2013ApJ...770..103H}) was designed and launched to study high-energy X-rays. It is the first telescope with the ability to focus on X-rays in the 3--79 keV energy range, providing a unique view of the universe. \nustar consists of two co-aligned telescopes with grazing incidence optics modules and focal plane modules, A ({\sl FPMA}) and B ({\sl FPMB}), consisting of a solid-state CdZnTe detector.

\begin{figure} \centering
    \includegraphics[width=\columnwidth]{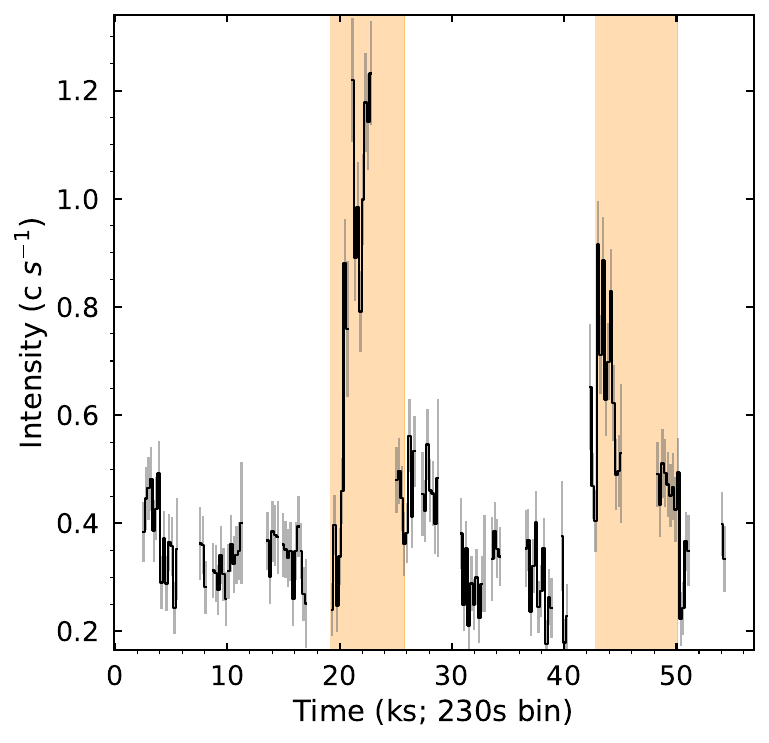}
    \caption{Background-corrected light curve of LS~I~+61~303 with a binning of 230~s, starting at 57979.9868 MJD. The rapid flares can be identified in the orange stripes. The errors in the count rate are shown in grey.}
    \label{Fig:lcb230}
\end{figure}

\nustar has technical capabilities that make it useful for a wide range of scientific studies. It has a high angular resolution, with a point-spread function (PSF) that varies from 18 arcsec at 10 keV to 58 arcsec at 79 keV. It can also measure spectral features with high efficiency \citep[see, e.g.,][]{Furst2014ApJ...784L..40F, saavedra, saavedra2023MNRAS}. These capabilities allow \nustar to accurately locate and study various types of astronomical objects. \nustar also has a good high-energy resolution, with a FWHM of 350 eV at 10 keV and 900 eV at 79 keV, which enables it to measure the energy of the incoming X-rays with unprecedented precision.

\nustar observed LS~I~+61~303 on 2017 August 14 (ObsID 90301008002), with an exposure of $\sim$ 55 ks. The data were already used in \cite{massi20} but not discussed in detail. The data were reduced using the {\tt NuSTARDAS-v. 2.0.0} analysis software from the {\tt HEASoft}~v.6.28 task package and {\tt CALDB} (V.1.0.2) calibration files. 
In order to filter the Southern Atlantic Anomaly passages, we looked at the individual observation report. 
We extracted cleaned event files using the parameters {\tt saacalc=1}, {\tt saamode=OPTIMIZED}, and {\tt tentacle=NO}. 
To extract the photons, we used circular regions of 70 arcsec centred at the source and similar nearby regions of 70 arcsec for the background, using the same chip. The chosen radius encloses $\sim90$ per cent of the PSF. The observation did not show stray light on any detector. The background subtraction of each camera module and the addition of corrected light curves were done using the {\sc LCMATH} task.
To generate the spectra, we used the {\tt nuproducts} task with the same regions as for light-curve extraction. The X-ray spectral analysis was performed using {\tt xspec} v.12.12.1~\citep{Arnaud1996ASPC..101...17A} in the 3--79 keV energy range. The spectra were grouped to a minimum of 20 counts per bin to properly use $\chi^2$ statistics.

 \begin{figure*} \centering
    \includegraphics[width=\columnwidth]{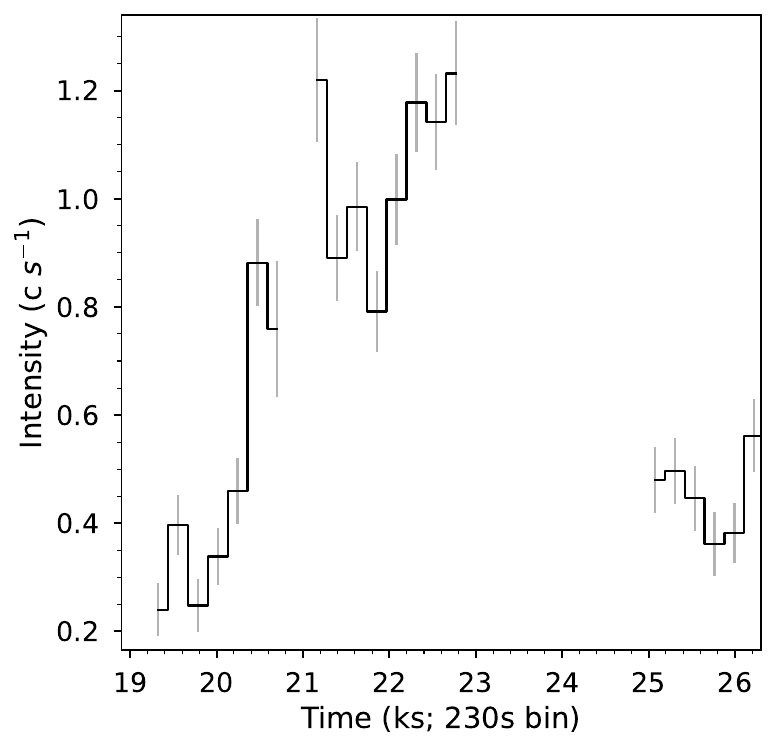}
    \includegraphics[width=\columnwidth]{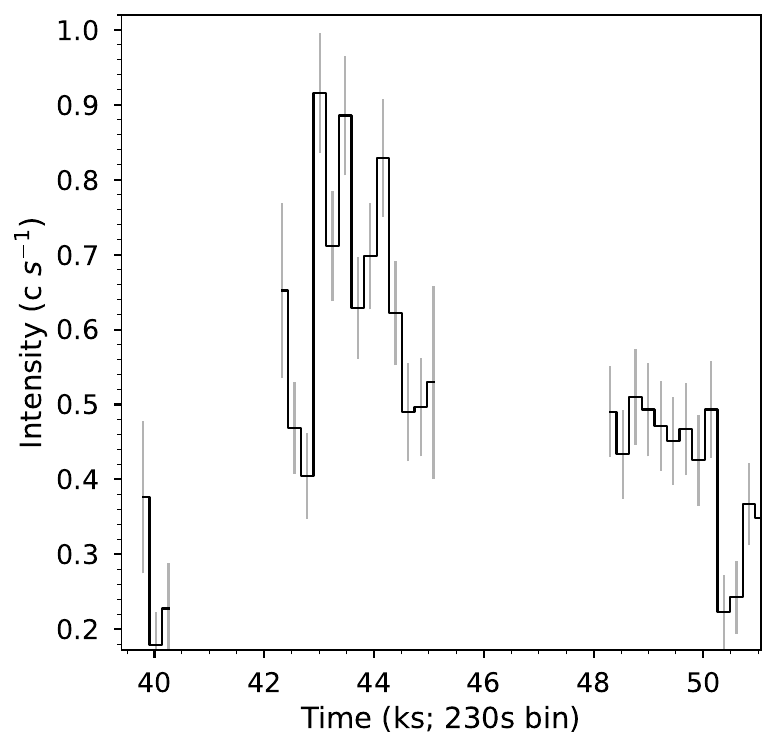}
    \caption{Light curve associated with each flare. Flare 1 is shown on the left, and Flare 2 is on the right.}
   \label{Fig:zoom}
\end{figure*}

 \begin{table}
 \centering
 \begin{tabular}{lccc}
 \hline
 Parameter & Units & Flare 1 & Flare 2 \\ \hline 
$L_{\nu}$ & $10^{33}$~erg~s$^{-1}$ & 10.5 & 8.4 \\
 $\Delta S$ & $10^{-11}$ erg s$^{-1}$cm$^{-2}$ & 1.17 & 0.81  \\
 $\delta t$ & ks & 5.9 & 7.9 \\
 $F_{v}$ & & 0.68 & 0.61 \\
 $\Gamma$ & & 2.05 $\pm$ 0.08 & 2.01 $\pm$ 0.10 \\
 $\chi^2$/d.o.f & & 88/102 & 81/71 \\
 \hline
 \end{tabular}
 \caption{Parameters of the two flares detected in the \textit{NuSTAR} data; $L_{\nu}$ is the luminosity assuming a distance of 2 kpc, $\Delta S$ is the amplitude flux density variation, $\delta t$ is the time-scale of the variation, $F_{v}$ is the variability fraction, $\Gamma$ is the power-law index , and $\chi^2$/d.o.f is the statistic associated with the fit. See the text for details.}
 \label{table:parameters}
 \end{table}

\section{Results}\label{sec:results}

In \hyperref[fig:bat]{Fig.~\ref{fig:bat}}, we show the orbital phase range of the \nustar observation in the context of the orbital light curve derived from \swift data. The \nustar observation spans in phase from 0.54 to 0.57, which is located before apastron passage ($\approx 0.7-0.8$), where high activity at different wavelengths has been observed in the past.

In \hyperref[Fig:lcb230]{Fig.~\ref{Fig:lcb230}}, we show the \nustar light curve with a time bin of 230 s. We identify two rapid flares that rise by a factor of $\sim 3$ compared to the persistent emission of the source. We estimate the variability time-scale as $\delta t=S_{\rm max} (\Delta S/\Delta t)^{-1}$, where $S$ is the count/rate associated with the light curve and $t$ is the time. The first flare had a variability time-scale $\delta t~\sim$ 5.9~ks, whereas the second one lasted for $\sim$ 7.9~ks. The amplitude flux density variation ($\Delta S$) for the first flare was $\approx 1.2\times 10^{-11}$ erg s$^{-1}$cm$^{-2}$, and for the second one it was $\approx 0.8 \times 10^{-11}$ erg s$^{-1}$~cm$^{-2}$. The variability fraction for each flare can be calculated using the formula \citep[e.g.,][]{Romero1995Ap&SS}

\begin{equation}
    F_{v} = \frac{S_{\rm max}-S_{\rm min}}{S_{\rm max}+S_{\rm min}},
\end{equation}    

 resulting in values of 0.68 for the first flare and 0.61 for the second one. These are the strongest flares reported in hard X-rays so far, except for the claimed magnetar activity reported by \citet{Dubus2008} using BAT data and later questioned by \citet{Rea2008ATel}, and the clear magnetar flare detected by \citet{Torres2012ApJ...744..106T} in the energy range 15--50 keV.

In \hyperref[Fig:zoom]{Fig.~\ref{Fig:zoom}}, we show a zoom into the flares. We see very fast rise times in their substructures, with count rate increases of a factor of $\sim 2$ between two contiguous bins. The decay in the first flare is unfortunately hidden by the Earth's transit. Thus, the decay time might be significantly shorter than what we have estimated. The second flare, in fact, shows a very fast decay on time-scales of $\sim 1$~ks. All this indicates significant changes in the hard X-ray emitter on scales of just several minutes, as seen in the past in {\it RXTE} data around apastron \citep{Smith2009ApJ}.

\begin{figure} \centering
    \includegraphics[width=\columnwidth]{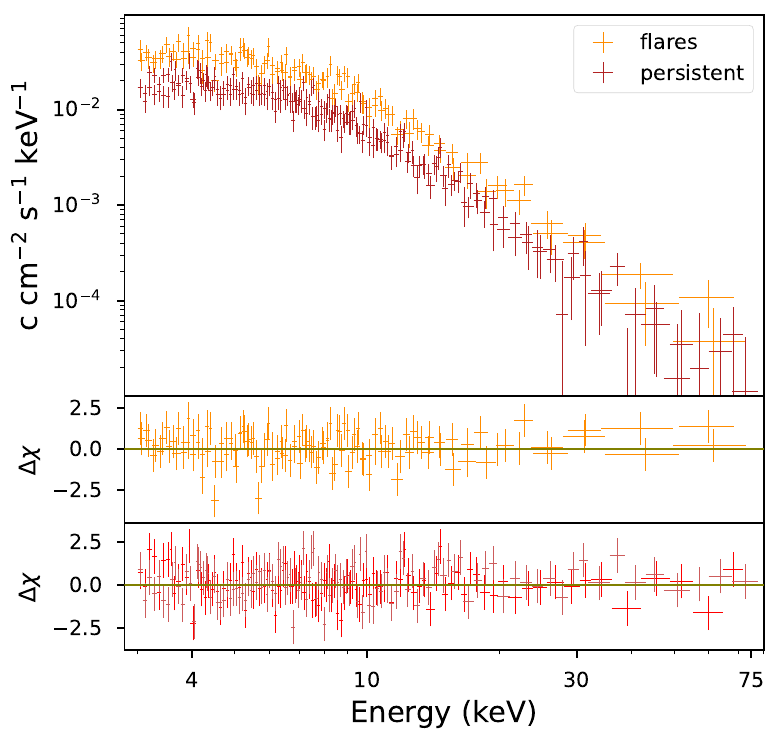}
    \caption{The energy spectra of LS~I~+61~303 observed by FPMA and FPMB show persistent emission and emission associated with the rapid flares that can be characterised by an absorbed power-law contribution in the energy range 3--79 keV.}
    \label{Fig:spectra}
\end{figure}

We obtained spectra of both the persistent emission and the flares. \hyperref[Fig:spectra]{Fig.~\ref{Fig:spectra}} shows the associated spectra. We fitted both spectra with an absorbed power law (in {\tt xspec}: {\tt tbabs*powerlaw}). The absorption column could not be correctly constrained at energies above 3~keV. Therefore, we fixed it at a value of 0.47$\times$$10^{22}$ cm$^{-2}$, which was derived from the \xmm observation analyzed by \citet{Chernyakova2017MNRAS.470.1718C} (observation X1). The orbital phase of the \xmm observation aligns with that of the \nustar observation. This model provided a very good fit to the data. For the average spectrum of the persistent emission, we obtained $\Gamma = 2.01 \pm 0.05$ with $\chi^2/{\rm d.o.f} = 115/125$. Upon observation, we found that the spectra of the flares are similar. We obtained $\Gamma = 2.05 \pm 0.08$ with $\chi^2/{\rm d.o.f} = 88/102$ for the first flare and $\Gamma = 2.03 \pm 0.10$ with $\chi^2/{\rm d.o.f} = 81/71$ for the second flare. We then computed the average spectrum for the flare activity, which had a value of $\Gamma = 2.03 \pm 0.10$, with a $\chi^2/{\rm d.o.f} = 198/225$, indicating that $\Gamma$ remains around 2, as in the quiescent state. The flares, then, are essentially \emph{achromatic}. 

We then obtained the flux associated with the {\tt cflux} model in the 3--79 keV range. If we assume that the source is located at $\sim$ 2 kpc, the luminosity associated with the persistent emission is $3.5~\pm~0.1\times$10$^{33}$ erg~s$^{-1}$, while the luminosity associated with the flares reaches $10 \pm 1.9~\times~10^{33}$~erg~s$^{-1}$ for the first one and $8.9 \pm 1.1 ~\times~10^{33}$~erg~s$^{-1}$ for the second one. The parameters obtained from the analyses described in this section are presented in \hyperref[table:parameters]{Table~\ref{table:parameters}}. 

\begin{figure} \centering
    \includegraphics[width=\columnwidth]{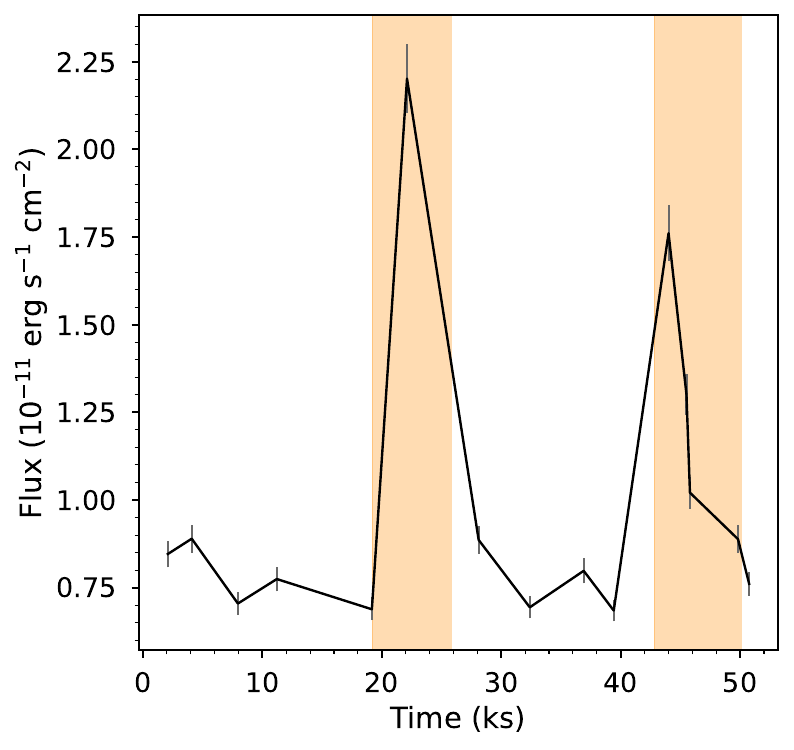}
    \caption{Light curve in terms of flux associated with each flare. To obtain the behaviour of the flares in relation to their dependence on flux, we partitioned the observation into different sections and derived the spectral fluxes. The orange stripes are associated with the flares, as shown in \hyperref[Fig:lcb230]{Fig.~\ref{Fig:lcb230}}.}
    \label{Fig:flux}
\end{figure}

In \hyperref[Fig:flux]{Fig.~\ref{Fig:flux}}, we show the flares expressed in flux units and over a wider orbital phase range, for a broader perspective. The solid line just connects the data points to facilitate the visual inspection. 
It is conceivable that we are missing relevant structures in the flares, but owing to gaps in the light curve, we are unable to determine this with certainty.  

Nevertheless, this does not hinder our ability, using the available information, to explore their nature and gain insights into the underlying mechanisms driving them.
 
\section{Discussion: origin of the flares}\label{sec:Discussion}

In this section we discuss qualitatively the physical origin of the flares found in hard X-rays. We will assume for the discussion that a powerful pulsar is present in the system, which is necessary to explain the high non-thermal luminosity of LS~I~+61~303. In this case, the X-rays are most likely of non-thermal origin, since the pulsar wind is expected to prevent any form of accretion (see, however, \citealt{Papitto2012ApJ...756..188P}). Moreover, no accretion features have ever been found in the X-ray spectrum of this source. In LS~I~+61~303 the non-thermal emission should come mainly from the shocked pulsar wind, which carries most of the available energy. Since \nustar can probe variability time-scales that are shorter than those probed at HE--VHE or in radio, due to the limited sensitivity of gamma-ray instruments and the typical size of the radio emitter, respectively, \nustar data offer us a unique opportunity to probe the physical processes behind the non-thermal emission in LS~I~+61~303.

Given the typical conditions in a high-mass binary, X-rays are likely to be of synchrotron origin \citep[see, e.g.,][]{tak09}, which implies that variability can be produced by changes in the magnetic field and associated losses, in addition to those due to particle acceleration or non-radiative/escape losses (i.e., adiabatic losses or heating and advection of the relativistic particles away from the binary). Flux variations induced by the changing Doppler enhancement caused by the reorientation of a relativistic emitting flow are also a potential source of variability. On the other hand, losses due to inverse Compton (IC) upscattering of stellar photons, the likely origin of the gamma rays \citep[see, e.g.,][]{bk09}, cannot change fast enough because the stellar photon field is steady and smoothly distributed in space\footnote{Note, however, that IC gamma-ray fluxes can still vary significantly if dominant synchrotron or non-radiative losses vary, or because of changes in electron acceleration.}. 

Since the variations found in hard X-rays are achromatic, the simplest mechanisms to explain the observed flares would be non-radiative (i.e., adiabatic or escape) losses or Doppler enhancement. Magnetic field variations that are faster than the cooling time of the emitting electrons could also be a possibility, leading to rapid variations in the synchrotron radiation. In fact, rapid magnetic field changes are naturally expected under flow evolution faster than radiative cooling, e.g. by compression, rarefaction, and turbulence, all on a time scale similar to those of adiabatic and escape processes (unless the field is decoupled from the plasma). There is some room, however, for synchrotron emission to dominate the energy losses, though still not determine the flare timescale, which will be discussed below. With all these mechanisms, the IC gamma rays should evolve similarly to the X-rays, although with different degrees of modulation. 

For electrons emitting at $\sim 70$~keV via synchrotron radiation and cooling by non-radiative losses in $\sim 10^3$~s, one can still conclude that the (potentially changing) magnetic fields should be (on average) less than $\sim 1$~G to keep the flare achromatic in the \nustar energy range (i.e., the synchrotron cooling time is kept $\gtrsim 1$~ks). On the other hand, the equipartition magnetic field in a mildly relativistic outflow is $\sim 10\,(L_{\rm pw}/10^{36}\,{\rm erg~s}^{-1})^{1/2}(d/10^{12}\,{\rm cm})$~G, where $d$ is the typical size of the emitting region ($10^{12}$~cm would be a plausible minimum), so $\lesssim 1$~G is expected within or near the binary system. The cooling time of IC with stellar photons in the Klein--Nishina regime for the same electrons is expected to be $\gtrsim 10^3$~s (as it should be if non-radiative losses dominate) for these $B$ values and any reasonable location of the emitting region. Despite all this, we think it worth mentioning a remaining opposite possibility in which $B$ is strong enough (say $\gtrsim 10$~G) for the electrons emitting from a few keV to $\sim 70$~keV to be cooled by the synchrotron. In this case, very rapid variations in $B$ may occur leading to very short flares, but so short that they would not be detectable, so that the observed changes in the emission could still be due to non-radiative processes such as those already described. Given the formula above, for such a $B$ strength the emitter should be located in the innermost regions of the two-wind interaction structure ($d\lesssim 10^{12}$~cm).

In the non-accreting pulsar scenario, electrons are accelerated in the regions where the relativistic pulsar wind and the stellar outflows interact. Arguably, the most natural variation time-scale of the emission in this scenario is associated with the angular velocity of the orbit, which at phase $\sim 0.6$ would yield variations on scales of $\sim 1$~day. However, shorter non-radiative variability is possible, and size constraints can be used to set lower limits in the variability time-scale. The smallest interaction region is located between the star and the pulsar, within the binary system (i.e., the intra-binary shock). In that region, the variability time-scale can in principle be as short as $\sim d/c$, which in LS~I~+61~303 is $\sim 10^2-10^3$~s; $d$ cannot be much smaller than the orbital separation distance ($d_{\rm orb}$) owing to the high pulsar-wind power. Interaction with a very dense Be disc may make $d$ significantly smaller. Such a situation is possible (as in the very high $B$ case mentioned), but would still require that the small emitter also changed in a time equal to the flare durations.
In principle, the evolution time in this region is in general set by the stellar-wind velocity, which is $\ll c$, but far quicker variations are still possible if the stellar wind is clumpy, as shocked clumps evolve much faster if they penetrate the unshocked pulsar-wind zone \citep{kb23}. We note that relativistic effects such as Doppler boosting should not play a very significant role in the intra-binary shock, because the shocked pulsar wind is weakly relativistic there \citep[where the flow is still subsonic; see, e.g.,][]{bog08}. 

\begin{figure} 
\centering
\includegraphics[width=\columnwidth]{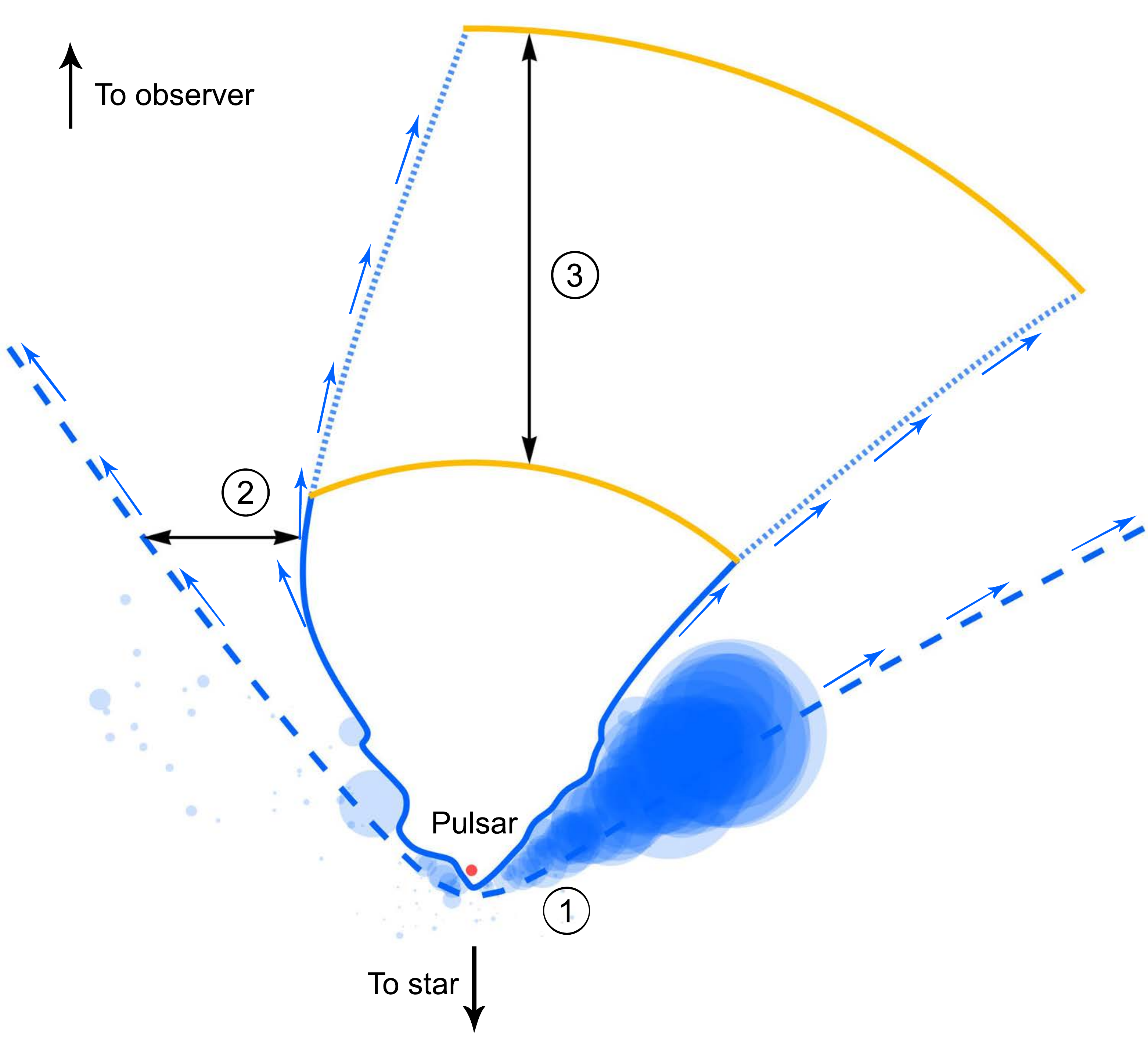}
    \caption{Three potential sub-hour variability mechanisms: changes in the intra-binary shock (thick solid blue line), e.g., because of clumps (translucent blue circles; 1); variations in the motion of the re-accelerated pulsar wind outside the binary with respect to the line of sight (blue and black arrows, respectively; 2); and strong fluctuations in the location of the Coriolis shock (yellow solid lines; 3).} 
    \label{Fig:sketches}
\end{figure}

Beyond the intra-binary shock, significant flux changes on scales of hundreds of seconds can be expected in a microblazar-like scenario, with the emitting flow suddenly changing direction \citep[e.g.,][]{rom02,Kaufman2002}. This can happen as the shocked pulsar wind becomes relativistic outside the binary \citep{bog08} and may suffer sudden re-orientations due to perturbations because of hydrodynamical instabilities, clumps, etc. \cite[e.g.,][]{zdz10,bbp15,hkr21,kb23}. These effects will be stronger when the pulsar is in the half of the orbit roughly before apastron, when the shocked flow can point toward the observer, as at the time of the \nustar observations \citep{Casares2005MNRAS.360.1105C,Aragona2009}. Another cause of fast variability outside the binary can be sudden changes in the rate at which particles escape or cool adiabatically in the emitting region downstream of the fluctuating shock that forms {\it behind} the pulsar \citep[the so-called Coriolis shock; see, e.g.,][]{bbp15,hkr21,kiss23}. Coriolis shock fluctuations are again caused by instabilities on spatial scales of $\sim d_{\rm orb}$ in the interaction structure. In both cases, the emitting flow should be at least mildly relativistic, which discards regions significantly affected by stellar-wind loading \citep{bbp15}. The variability time-scale of fluctuations in the Coriolis shock is thus $\sim d_{\rm orb}/c$, yielding sub-hour time-scales in LS~I~+61~303. Doppler boosting variability is determined by the dynamical time-scale of the perturbed shocked-pulsar-wind region that produces velocity direction changes further downstream, and so it is $\sim d_{\rm orb}/c$ as well. To illustrate the qualitative picture presented here, in \hyperref[Fig:sketches]{Fig.~\ref{Fig:sketches}} we show a sketch of the potential variability mechanisms that may explain the hard X-ray fast changes. Summarising, these are: 
\begin{itemize}
\item Changes in the intra-binary shock due to stellar-wind perturbations, such as clumps (translucent blue) leading to significant variations in the emitter size and thus to non-radiative losses (see 1 in the figure). 
\item Variations in the flow direction (and velocity) of the re-accelerated pulsar wind outside the binary. The blue arrows give the directions of the flow with respect to the observer direction, which determine Doppler boosting effects (2).
\item Strong fluctuations in the location of the Coriolis shock because of changes in the conditions in the region where such a shock is triggered, that is, the side of the two-wind interacting structure facing the stellar-wind lateral impact due to orbital motion (3).
\end{itemize}

 We note that the likely variations in the magnetic field associated with the changing flow properties will also contribute to the overall radiation changes, but the field must not be dominant in modulating the synchrotron emission via losses for the reasons given above.

We also note that the location and size of the clumps in \hyperref[Fig:sketches]{Fig.~\ref{Fig:sketches}} and the shape of the contact discontinuity affected by the clumps have been calculated as explained in \cite{kb23}, while the curvature of the contact discontinuity outside the binary and the approximate location of the Coriolis shock roughly follow the simulation results \citep[e.g.,][]{bbp15,hkr21}.

It has already been proposed in the literature that the stellar-wind structure may be responsible for the variations in these systems (although as already mentioned stellar-wind influence should be indirect, as mass loading should be small in the X-ray-emitting region). The unstable shocked pulsar wind, Doppler boosting effects, or a combination of all these effects, have also been contemplated in the past, and we consider these processes as the most likely explanations for the behaviour found here. 

It is worth mentioning that other gamma-ray binaries also present short-term variability in their X-ray emission. For instance, LS~5039, a gamma-ray binary roughly similar to LS~I~+61~303, also features sub-hour variations in the X-ray light curve (as discussed, e.g., in \citealt{bos05}, and most recently in \citealt{yon23}; see also \citealt{vol21} for hints of even faster variations). The gamma-ray binary 1FGL~J1018.6-5856 could be another related example \citep{an15}. Therefore, the conclusion that the X-ray emitter should be located close to or within the binary, and made of relativistic plasma, may be generalised to other gamma-ray binaries. Efficient wind--wind mixing is expected from simulations \citep[e.g.,][]{bbp15}, so the X-ray-emitting zone may be restricted to the region encompassing the intra-binary shock and up to the immediate vicinity of the Coriolis shock (but close to the two-wind contact discontinuity, where mixing is the strongest). To finish, we note that strong fast VHE variability in these sources is hinted by the detection of sub-day (or shorter) scale flares \citep[e.g.,][]{aha06,ver16}. If confirmed, this would indicate that the X-ray and VHE emitters largely overlap unless $B$ is too high \citep{Zabalza2011A&A...527A...9Z}.

\section{Conclusions}\label{sec:Conclusiones}

We have identified two fast achromatic flares in the hard X-ray (3--79 keV) light curve of LS~I~+61~303 before the apastron passage of the compact object. These flares are very rapid, with time-scales of a few ks for the whole flare, and with intra-flaring variability times of a few hundred seconds. The spectrum and the expected physical conditions of the emitter strongly suggest a synchrotron origin for the radiation. Moreover, the achromaticity points to non-radiative losses (either escape or adiabatic cooling/ heating) in a relativistic flow as the dominant processes in the electron evolution, and the rapid flux changes are most naturally explained by sudden changes in the size of the emitting region and/or its motion with respect to the line of sight, with fast magnetic-field variations tied to the flow and Doppler boosting possibly playing an important role. A rather high $B$-field cannot be discarded, although the X-ray variations are more naturally explained by flow processes. The sudden changes of the emitter can be caused by various situations, such as changes in the intra-binary shock (e.g. due to clumping), variations in the re-accelerated pulsar wind outside the binary, or strong fluctuations in the location and size of the Coriolis shock region. Future multi-wavelength observations could be useful to constrain the physical processes behind the rapid hard X-ray flares in LS~I~61~303; in particular, to determine the clumpy nature of the stellar wind and the details of the non-thermal emitter structure and dynamics. The results of this study may also have implications for other gamma-ray binaries with similar X-ray variability.

\section*{Acknowledgements}
E.A.S. is a fellow of the Consejo Interuniversitario Nacional, Argentina. G.E.R., V.B-R., and E.K. acknowledge financial support from the State Agency for Research of the Spanish Ministry of Science and Innovation under grant PID2019-105510GB-C31AEI/10.13039/501100011033/ 
and through the ''Unit of Excellence Mar\'ia de Maeztu 2020-2023'' award to the Institute of Cosmos Sciences (CEX2019-000918-M). Additional support came from PIP 0554 (CONICET). V.B-R. is Correspondent Researcher of CONICET, Argentina, at the IAR.

\section*{Data Availability} 
This research has made use of data obtained from the High Energy Astrophysics Science Archive Research Center (HEASARC), provided by NASA’s Goddard Space Flight Center.


\bibliographystyle{mnras}
\bibliography{biblio}







\bsp	
\label{lastpage}
\end{document}